# PROPERTIES OF THE X-RAY PULSAR GX 301-2 IN HARD X-RAYS[a]

V. Borkus[1], A. Kaniovsky[1], R. Sunyaev[1], V. Efremov[1],
P. Kretschmar[2], R. Staubert[2], J. Englhauser[3], W. Pietsch[3]

[1] *Russian Space Research Institute, RAS, Moscow*
[2] *Astr. Inst. der Univ. Tuebingen, Waldhauserstr. 64, D-72076 Tuebingen, Germany*
[3] *Max-Planck-Institut für extraterrestrische Physik, D-85470, Garching, Germany*

In 1993-1994 a series of observations of the X-ray pulsar GX 301-2 by HEXE onboard Mir-Kvant was made. A period of pulsations was measured (it varied between 675 and 678 s) and pulse profiles in different energy bands were produced. The measured luminosity in the 20-100 keV energy range changed substantially between $8 \times 10^{34}$ and $7 \times 10^{35}$ $d^2$ erg/s (d is the distance to the source in kpc). The obtained spectrum is quite satisfactory described by the "canonical model for X-ray pulsars" with $\gamma=1.3$, $E_c \sim 23$ keV, $E_f \sim 9$ keV. It changed weakly between the observations, but was softest at brightness maximum. Significant variations of the spectral hardness over the pulse phase were detected, but the accumulated data are insufficient to quantify variations in spectral parameters. No significant traces of cyclotron lines were found. An interpretation of the pulse profiles as superposition of emissions from two flat polar caps (with inclusion of gravitational lensing) leads to an estimate of the angle between the magnetic axis and axis of rotation of 40-70° and an angle between the direction to the observer and the rotation axis of 75-85°.

## INTRODUCTION

The X-ray source GX 301-2 (4U1223-62) has one of the longest period among the known X-ray pulsars. Together with the blue supergiant Wray 977 of spectral class B1.5Ia it forms a binary system, which has an orbital period of $41^d.508$ and an eccentricity of 0.472 (Sato et al, 1986). Parameters of the optical star (radius $\sim 43 R_\odot$ and mass $\sim 30 M_\odot$) and the distance to the source ($d \sim 1.8$ kpc), thought to be known for almost 20 years (Vidal, 1973; Johes et al., 1974; Parkes et al., 1980) are currently questioned by recent observations (Kaper et al, 1995). The newer data indicate a distance of 5.3 kpc. On the other hand, Koh et al. (1997) argue that at least $d<4.3$ kpc.

When the pulsar approaches the optical companion, the X-ray luminosity of the binary system grows and reaches its maximum $1^d.2$ before the periastron passage (White and Swank, 1984; Stivens, 1988; Chichkov et al. 1995; Koh et al., 1997), i.e. at orbital phase $\varphi_{orb} \sim 0.96$. The lower maximum appears also at $\varphi_{orb} \sim 0.5$.

The orbital light curve near a maximum is asymmetrical: the source flux smoothly grows to the maximum and then drops fast. This effect was explainned by Krasnobayev and Sunyaev (1977) and Chichkov et al. (1995) as follows. When the accretion rate reaches a certain critical value, the pulsar luminosity becomes sufficient to warm up a star wind. A standing shock wave is formed, and the accretion is reduced.

The strong star wind heavily absorbs X-rays and reemits them in a strong fluorescent iron line. The surface hydrogen column ($N_H$) in the source direction can reach $10^{24}$ cm$^{-2}$, according to Leahy and Matsuoka (1990). The same data show that $N_H$ and the iron line intensity are correlated, indicating that both peculiarities are caused by interaction of X-ray radiation with the star wind.

---

[a] To be published in the Soviet Astronomy Letters, 1998

Since 1984 the pulse period has decreased from ~700 s to ~675 s. The smallest known value of the period (675.7 s) was detected in July 1994 (Pravdo et al., 1995). The BATSE observations in 1991-1995 revealed that currently the period oscillates around an average value of ~678 s (Koh et al., 1997).

The behaviour of the source in a standard X-ray band 1-20 keV is well investigated (Leahy and Matsuoka, 1990; Pravdo et al. 1995; Haberl, 1991), but the information about its properties in the hard (> 20 keV) energy band is still poor.

OBSERVATIONS

HEXE is a phoswich detector system consisting of four units having a total collecting area of about 800 cm$^2$. The crystals in each detector are 3.2 mm NaI(Tl) backed by 50 mm CsI(Tl). The field of view is restricted by a tungsten collimator to 1.6x1.6° FWHM. The instrument allows to measure the source spectra in the energy range 20-200 keV with 30% resolution at 60 keV. The observations were performed in a collimator rocking mode (2 detectors on source and 2 detectors on background with 2.3° offset). The duration of on source and background pointings are ~120 s. The low orbit of the Mir station does not allow continuous monitoring of the source and the observations are done in "sessions". The duration of a typical session is between 7 and 30 minutes.

To correct the data for collimator efficiency the exact orientation of the Mir station has to be known. At present it is determined from the simultaneous observations by the TTM telescope also installed onboard Mir-Kvant. The accuracy of the pointing determination is several arc minutes. The calibrations of the third and fourth detector show that their physical characteristics differ substantially from those of the first pair. In particular they have worse energy resolution. So for source spectroscopy we used the first two detectors only; the third and the fourth detectors were used in timing and pulse profile analysis. A more detailed instrument description may be found in Reppin et al. (1983).

In 1993-94 four sets of observations of GX 301-2 by Mir-Kvant were made. Information about these observations is summarized in Table 1.

TIMING

The periods determined using the epoch folding technique by Leahy et al. (1983) are given in Table 1. The pulse profiles are presented on Fig. 1-2. We note, that for different observations we chose different phase bin sizes. In the first and the fourth set the bin size was 1/60; during the second set the source was in a low state, and to improve the statistic in the pulse profiles we used bins with size 1/30; in the third set the observational time was small, and we have taken the phase bin size 1/40.

We corrected the doppler shifts due to the instrument motion in respect to sun and motion in the binary system using the ephemeris of Sato et al (1986). The orbital parameters of the binary system are listed in Table 2.

Table 1
Observations

| Set # | Date | JD 2440000+ | On-source time, s | Period, s | $L_X$, $10^{35}$ d$^2$ erg/s, 20-100 keV[*] | Orbital phase |
|---|---|---|---|---|---|---|
| 1 | 3-10.1.93 | 8991-8998 | 15400 | 677.61±0.01 | 2.7±0.3 | 0.49-0.66 |
| 2 | 4-9.1.94 | 9357-9362 | 4500 | 676.35±0.1 | 0.85±0.4 | 0.31-0.43 |
| 3 | 2.6.94 | 9505.45-9505.6 | 1846 | 676.2±1.9 | 6.8±0.8 | 0.88-0.89 |
| 4 | 6-9.6.94 | 9510.4-9512.5 | 3745 | 675.7±0.1 | 2.2±0.4 | 0.007-0.06 |

[*] $d$ is the distance to the source in kpc

In all cases the source was detected up to ~100 keV, and in the first set of observations the pulsations were detected also in the hard energy band (48-75 keV).

The amplitude of pulsations (rms), defined as a root-mean-square fluctuation of the source count rate over the pulse, has appeared to be the same (~42 %) within uncertainties in sets 1, 2, 4, but much lower (~25%) in the third set, where the source was in the brightest state; rms also grows slightly towards the higher energies, and is ~50% in the 30-50 keV energy range.

Table 2
Binary system parameters[1]

| | |
|---|---|
| $P_{orb}$ | $41^d.508$ |
| e | 0.472 |
| $a_x \sin i$, lt. sec. | 371.2 |
| $\tau$, JD 2440000+ | 3890.406 |
| $\varpi$ | $-50°.1$ |

[1] see Sato et al (1986)

The rms reduction when the source was in the brightest state can be explained in several ways. For example, during the burst a significant amount of matter can concentrate near the compact object. This matter can intercept and scatter X-ray radiation. It is also possible that the structure of the polar column substantially changes when the luminosity grows.

In two of four cases (sets 1, 4), when the accumulated statistics was high, the strong variations of the source hardness (defined as a ratio of the source fluxes in the energy bands 30-48 keV and 14-30 keV) over the pulse phase were found, assuming that the pulse profile depends on the energy range. The average hardness is ~0.3. The strongest deviations were observed January 3-10, 1993 (set 1); the probability of their casual occurrence is $2 \times 10^{-17}$ ($\chi^2$=201/59). The hardness minima occur at the pulse phases 0.5-0.7 and 0-0.2. It is notable, that in set 3 the source hardness differed from the standard being it was 1.3 times lower (~0.23).

The source power density spectrum produced using the data from sets 1,2, 4 is shown on Fig. 3. It is roughly described by a power-law $P(f)=P(0.02\ Hz)(f/0.02\ Hz)^{-\gamma}$ with slope $\gamma=1.9\pm0.5$ and normalization $P(0.02\ Hz)=1.11\pm0.29\ Hz^{-1}$. The integral noise power in the 0.02-5 Hz frequency range is rms=25.2±5.0%; the most significant input to this noise (~90%) is produced by variations with frequencies 0.02-0.5 Hz.

The observations of pulses, accumulated for individual sessions show that each of them can differ from the average one. This effect can happen due to a number of reasons: for example, due to variability of the accretion rate on time scales smaller than the period of pulsations.

SPECTROSCOPY

The source spectrum is shown in Fig. 4. It is well described by the model "canonical for X-ray pulsars" (White et al., 1983) with fixed value of photon index $\gamma=1.3$:

$$I_c(E) = I_0 E^{-\gamma} \begin{cases} 1, & E < E_c \\ \exp(-(E-E_c)/E_f), & E > E_c \end{cases} . (1).$$

The HEXE energy range is insufficient for independent measurements of $\gamma$, therefore we did not try to determine this parameter from our data, and have taken it from simultaneous observations of the source by the TTM telescope (Aleksandrovich et al. 1995). The results obtained using this approximation as well as the thin thermal bresstrahlung fit results are presented in Table 3.

The spectra obtained in three sets of observations are identical, within uncertainties. The spectrum during the high state (2.4.94) is slightly softer than average (see fig. 5). The position of a break characterised by $E_c$ remained almost constant at ~23 keV during the observations. Variations of parameter $E_f$ are hardly more essential: $E_f$ is consistent with constant with $\chi^2$=6.3/3 (probability ~10%).

No significant traces of cyclotron lines were found in any of the spectra.

To investigate the spectral variations over the pulse phase we have carried out spectra accumulation in four phase intervals: 0.95-0.18, 0-0.18-0.45, 0.45-0.67, 0.67-0.95, which correspond to the two minima and two maxima of the pulse profile. Then they were fitted by the "canonical" model. The results for one of the observations are shown in Fig. 6. It is clear, that within uncertainties, the spectral parameters are consistent with a constant value, indicating insufficiency of the accumulated statistics for more detailed analysis than based on measurements of the hardness ratio. In general, the spectral data are well consistent with the results of observations by HEAO A-1 (Rothshild, Sung, 1986) and with measurements of hardness ratio, described in section Timing.

DISCUSSION

According to Parkes et al. (1980) the distance to GX 301-2 is 1.8 kpc. Using this value and extrapolating the spectrum measured by HEXE to the energy range 1-100 keV, we receive source luminosities for the different observations between $9 \times 10^{35}$ erg/s and $8 \times 10^{36}$ erg/s. This is about three times higher than in the range 20-100 keV. Gravitational redshift can reduce the luminosity $1/\sqrt{1 - R_g/R_{ns}} \leq 1.6$ times ($R_{ns}$, $R_g$ are the geometrical and gravitational NS radii, $R_{ns} > 1.6 R_g$), so the above estimates must be slightly increased. Consideration of radiation beaming can slightly reduce the luminosity estimate (no more than 2 times).

According to Basko and Sunyaev (1976) there is a critical luminosity

$$L^* = 4 \times 10^{36} \left(\frac{\sigma_T}{\sigma}\right)\left(\frac{l_0}{2 \times 10^5 \text{cm}}\right)\left(\frac{10^6 \text{cm}}{R}\right)\left(\frac{M}{M_\oplus}\right) \text{ erg/s} \quad (2)$$

($\sigma$ is a total scattering cross-section of matter in the accretion column, $\sigma_T$ is the Thomson cross-section, $l_0$ is an accretion column length in an azimuthal direction, $R$ is NS radius, $M$ is its mass) that separates two accretion regimes. In the first regime the influence of radiation on the accreting matter is negligible. In the second one this influence is significant. When $L \leq L^*$ the free-fall zone goes almost to the NS surface (Basko & Sunyaev, 1976), and the polar cap emits preferably in directions normal to this surface. A characteristic angular width of a pencil beam is ~15° (Basko & Sunyaev, 1975).

Table 3.
Spectra hardness and pulse fraction

| Set | hardness * | rms(14-30 keV), % | rms(30-48 keV), |
|---|---|---|---|
| 1 | 33.2±0.3 | 42.6±0.5 | 48.2±1.0 |
| 2 | 30.5±2.4 | 46.1±4.4 | 56.3±7.7 |
| 3 | 25.6±0.7 | 24.8±0.9 | 28.4±2.0 |
| 4 | 31.9±0.9 | 40.0±1.2 | 52.4±2.6 |

* Ratio of the source fluxes in the 30-48 keV and 14-30 keV bands

Table 4. The results of the spectral fit by the "canonical model for X-ray pulsars"

| Set | γ* | $E_c$, keV | $E_f$, keV | $\chi^2$/dof | $kT_{bremm}$ |
|---|---|---|---|---|---|
| 1 | 1.3 | 23.8±1.2 | 8.6±0.5 | 7/9 | 11.5±2.5 |
| 2 | 1.3 | $21.4_{-48}^{+5}$ | 11.5±3 | 10.5/10 | 13.5±2 |
| 3 | 1.3 | 23.3±1.2 | 7±0.5 | 6.6/10 | 9.6±0.23 |
| 4 | 1.3 | 22.7±2 | 8.7±1.2 | 5.6/9 | 11.4±0.5 |

* Photon index is fixed ar γ=1.3

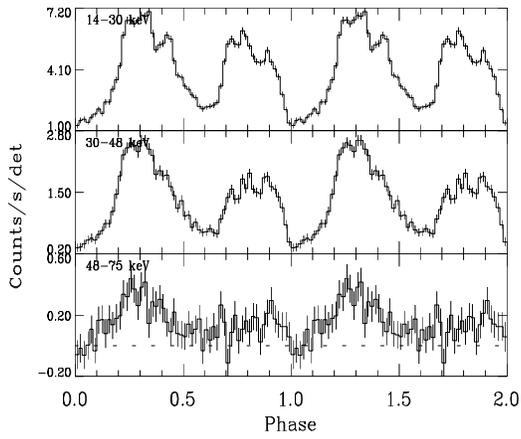

3-10.1.93

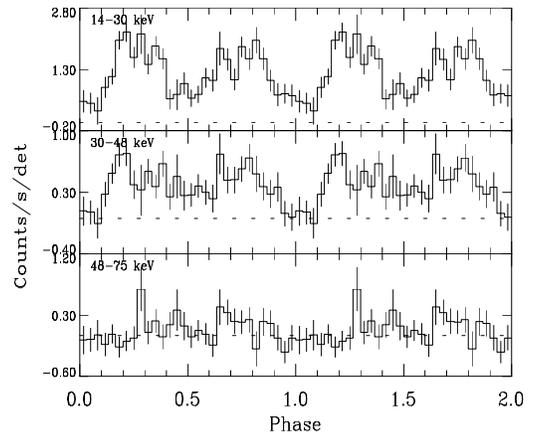

4-9.1.94

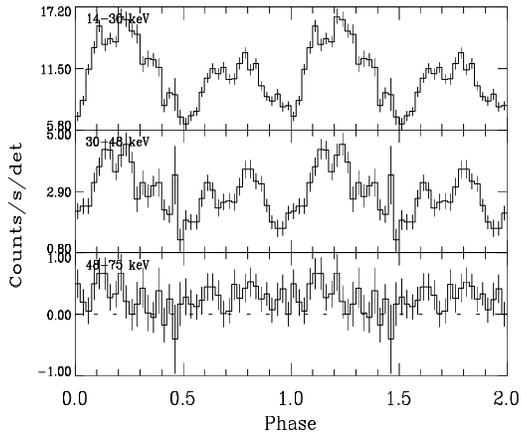

2.6.94

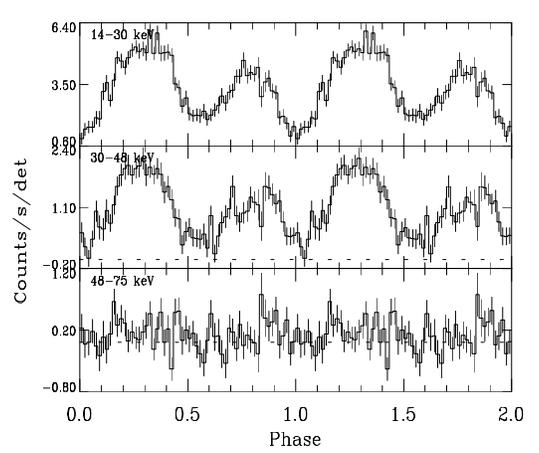

6-9.6.94

Fig. 1 The pulse profiles of GX 301-2 produced during the observations made 3-10.1.93 (orbital phase $\varphi_{orb}$=0.49-0.66), 4-9.1.94 ($\varphi_{orb}$=0.31-0.43), 2.6.94 ($\varphi_{orb}$= 0.887-0.89) and 6-9.6.94 ($\varphi_{orb}$=0.006-0.56).

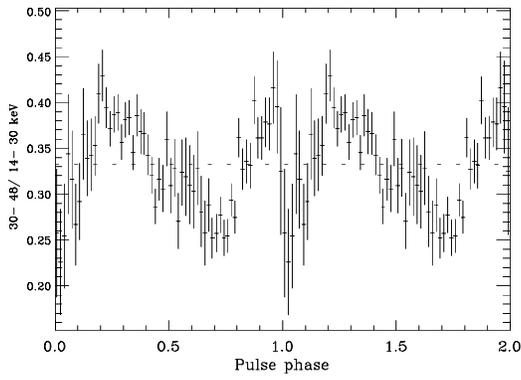

3-10.1.93

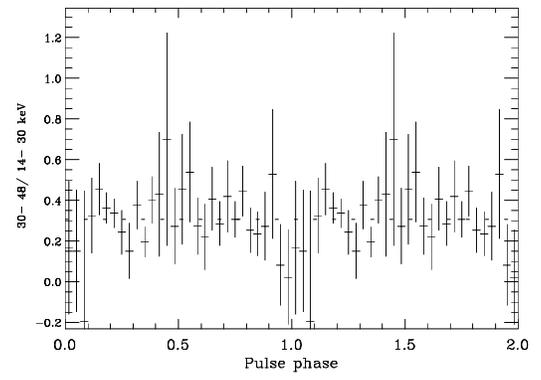

4-9.1.94

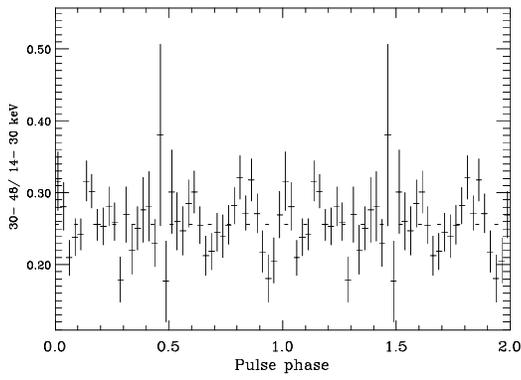

2.6.94

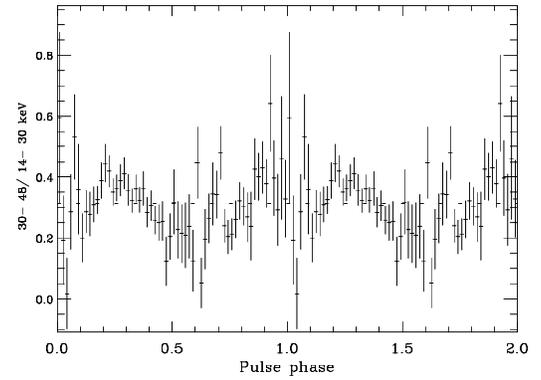

6-9.6.94

Fig. 2. The hardness ratio p (defined as Flux(30-48 keV)/Flux(14-30 keV)) changes over the pulse phase. The data correspond to 3-10.1.93 (p=0.31, $\chi^2$=201/59), 4-9.1.94 (p=0.3, $\chi^2$=21.7/29), 2.6.94 (p=0.228, $\chi^2$=85.1/59), 6-9.6.94 (p=0.286, $\chi^2$=97.1/59).

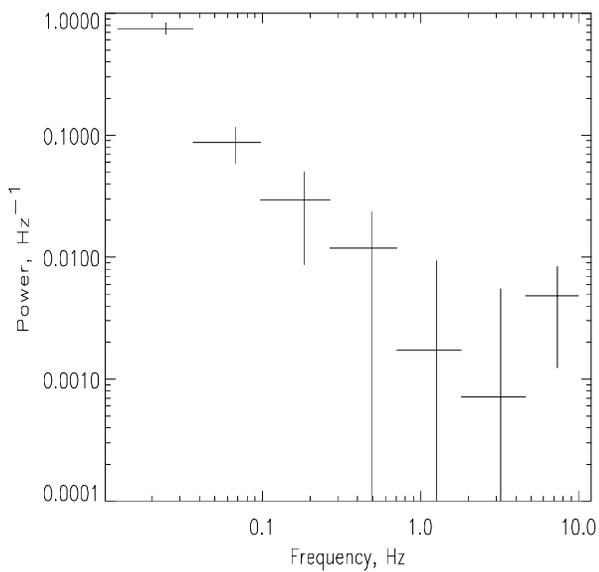

Fig. 3. The power density spectrum of GX 301-2 obtained in the 20-50 keV energy band.

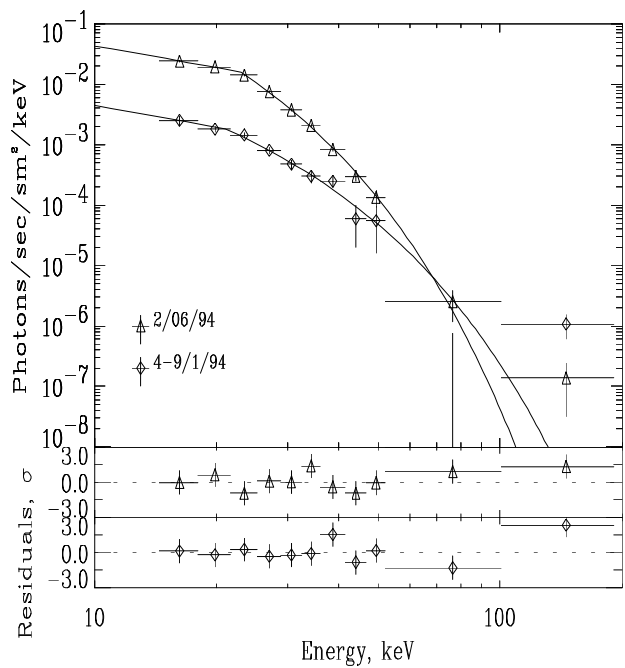

Fig. 4. The observed spectrum of GX 301-2 at the brightness minimum (4-9.1.94) and maximum (2.6.94).

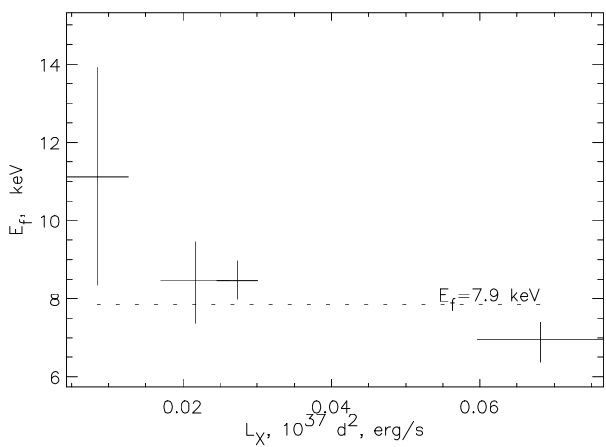

Fig. 5. Relation between the energy $E_f$ and the luminosity $L_X$ (in the energy range 20-100 keV). Test for consistency with constant $E_f$=7.9 keV gives $\chi^2$=6.3/3

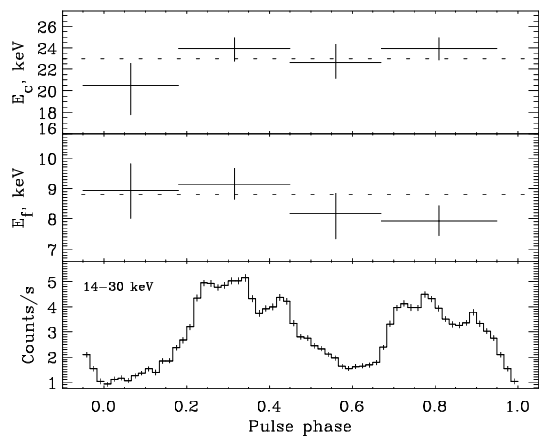

Fig. 6. Dependence of spectral parameters on the pulse phase for observations 3-10.1.93. Both parameters are in good agreement with their averages. See text for the definition of phase intervals.

The intensity distribution over the viewing angle is roughly described by
$$I(\mu) \propto \cos^n(\mu) \quad (3),$$
with n between 1-3, at all energies, except those close to a cyclotron resonance (Meszaros et al, 1985). The angle $\mu$ is counted from normal to the NS surface at a particular point of the polar cap. The parameter *n* determines beaming of radiation; the higher the *n*, the stronger the source flux varies over the pulse phase. Our data indicate that during all the observations, except that of April 2 1994, the pulsar luminosity was lower than $L^*$. (We shall note in connection with this, that during the April 2 1994 observations the hardness of the spectrum and rms were different than in the other cases.)

In recently published work by Kaper et al. (1995) another estimate of the distance to the source ($d\sim5.3$ kpc) is given, based on comparison of Wray 977 optical spectrum with the spectrum of a supergiant $\zeta^1$ Sco. If this estimate is correct then the pulsar luminosity will be an order of magnitude higher ($L_X\sim10^{37}$ erg/s), but still will be sufficiently small to allow us to use the hot spot simplification. Whereas, in the brightness maximum, the luminosity will be $\sim10^{38}$ erg/s, and two accretion columns will be formed at the poles, extending upward along the magnetic field lines and radiating mainly sideways. The source spectrum in the latter case will be appreciably softer than in the low state, and rms will be lower (this, by the way, is observed). In addition, the beaming of radiation will change, because the column will radiate sideward, and not upward. Hence, the pulse phase will shift by 0.25. However, comparison of two observations made in June 1994 shows that the detected phase lag of $\sim0.1$ is much smaller than expected, and is consistent with that caused by inaccuracy of period determination ($\sim0.08$). As it was already mentioned, the results by Kaper et al. (1995) are inconsistent with other optical observations (see Koh et al., 1997). So, we will use the 1.8 kpc estimate for further analysis.

EMISSION MODELLING

Various models of NS structure point at a NS radius of $\sim2$-3 $R_g$. Therefore, bending of photons trajectories in the pulsar gravitational field will strongly influence the structure of the pulse profiles. According to Pechenick et al. (1983) and Riffert and Meszaros (1987) the angular dependence of radiation of the polar cap can be written as
$$I(\theta_0) \propto \int_0^1 f(\mu(x)) h(x,\alpha,\theta_0) x\, dx \quad (4),$$
where $x=\sin\mu$, $f(x)$ is the distribution given by equation 3, and the function h, which describes the geometry of a radiating surface, is determined by formulas:
$$B = \frac{\cos\alpha - \cos\theta_0 \cos K}{\sin\theta_0 \sin K} \quad (5)$$

$$h(\theta,\alpha,\theta_0) = \begin{cases} 0, & B>1,\ K<\theta_0-\alpha,\ K>\theta_0+\alpha \\ \pi, & B<-1 \\ \arccos(B), & -1<B<1, \end{cases} \quad (6)$$
where
$$K = x\int_0^1 ((1-1/R) - (1-\rho/R)x^2\rho^2)^{-1/2} d\rho \quad (7)$$

is the angle between vectors from the NS center to the emission point and direction to observer (in equation (7) radius $R$ is expressed in terms of $R_g$), $\alpha$ is the angular size of the polar cap, $\theta_0$ is the angle between the vector from the NS center to the polar cap center and the direction to observer. Including the effects of the gravitational field would make the pulse profiles less sharp.

We have used the above formalism for a description of the pulse profile and determination of the NS parameters.

An experimentally found pulsar acceleration $\dot{P}/P \approx -8.8 \times 10^{-3}$ year$^{-1}$ (see, for example, Lutovinov et al. 1994) requires the presence of an accretion disk. Therefore, the polar column should be empty inside, and the cap formed in its basis should have a ring form. For numerical calculations we have assumed that the angular radius of this ring is $\alpha=6°$ and its thickness is $\Delta\alpha=0.1\alpha$, i.e. they are equal to values, received from the Ghosh & Lamb (1978, 1979a, b) accretion theory. The ring emission was simulated as the difference of the emissions of two solid caps of appropriate sizes. The choice of exact values of $\alpha$ and $\Delta\alpha$ is not very significant: subsequently, varying these parameters, we were convinced, that their small changes result in insignificant changes of the general picture.

For example, if the accreting matter does not form the disk, but is being caught directly into the accretion column, then the polar cap should be solid. However, fitting by the models with solid and empty caps yield similar results. The difference becomes prominent only when the angle between the magnetic axis and the axis of rotation is greater than 85°.

Presence of pulsations means, that the magnetic axis is inclined in respect to the axis of rotation and, probably, that of the accretion disk. Therefore the angular extent of the column (and the radiating surface) in the azimuthal direction is less than 360°. We shall note, that equation (4)-(7) were received under the assumption of ring symmetry, and consideration of any other form of the emitting surface requires updating of function h, which becomes dependent on the pulse phase. In this work both the simplified version of the geometry (in which the extent of the column is 360°) and the more complex geometry (where the extent is 120°) are considered.

Free parameters of our model are the angle $\beta$ between the magnetic axis and the axis of rotation, angle $\Gamma$ between the direction to the observer and the axis of rotation, the initial phase $\phi_0$, the azimuthal shift of the second pole in respect to a diametrically opposite position $\Delta\phi$ (counted in the direction of rotation), a shift $\Delta\beta$ of this pole toward the plane of rotation, and the parameter $n$ describing radiation beaming. One more variable is the pulse amplitude, which is fitted by the least squares method for each set of the specified parameters. We note, that taking gravitational effects into account results in easing the correlation between $\beta$ and $\Gamma$, which is inevitable in a nonrelativistic case (Riffert et al., 1993).

For approximation we used the 14-30 keV pulse profile, obtained January 3-10, 1993, in which statistics was best. Fitting results are presented in Fig. 7 and in Tables 5,6. Model agrees with the experimental data with accuracy ~20%, so it is possible to talk only about a qualitative description of the pulse shape.

For analysis we used the threshold $\chi^2=1000$ which corresponds to a root mean square deviation of ~25%. All fits with $\chi^2>1000$ where considered as bad and where not included in Table 5. In this area fall all the models, which suppose isotropy of the polar cap emission, therefore they can be reliably excluded from consideration. In observations, in which the accumulated statistics was lower (sets 2 and 4), the model is consistent with the data with $\chi^2$ per degree of freedom smaller than 1.5.

It follows from Table 5, that the results are very dependent on the beaming factor. The model allows to exclude all values of n, which are smaller than 2. However, varying n from 2 to 3 gives very different values of $\beta$ while maintaining the quality of approximation. The choice of geometry of the Alfven surface also significantly influences the estimation of the angle β. However, to determine its shape it is necessary to solve an appropriate magnitohydrodynamic problem, and is beyond the scope of this paper. It has been seen from Table 5 that if $n$~2.5-3, the magnetic axis should be inclined to the axis of rotation by the angle of 40-70°.

The model indicates that the angle Γ under which the observer looks at the NS (in respect to the axis of rotation) is ~75-85°. According to Sato et al. (1986) the absence of eclipses means, that the inclination angle of the binary system is $i<78°$ (on the basis of expected optical star mass, $i$~75° is proposed). This means, that the pulsar rotation plane is inclined to the binary plane by an angle smaller than 10°. We shall note here, that the reconsideration of the optical star parameters (spectral class B1 Ia +, mass $M>48\ M_\odot$ -- see. Kaper et al., 1995) leads to another estimate of the binary system inclination ($i<64°$).

The difference in the pulse minima depth means that the poles are not diametrically opposite, but are azimuthally shifted on $\Delta\phi$~-10°, i.e. the pulsar magnetic field obviously can not be described by the dipole approximation. Besides the azimuthal shift, the second pole can be displaced in the other direction: by an angle $\Delta\beta$~10-20° towards the plane of rotation. In the current model this angle is correlated with Γ (with the dependence $\Gamma\approx\Gamma(\Delta\beta=0)-\Delta\beta/2$). For the specified parameter range the uncertainty in Γ is about 5°.

The NS radius is correlated with the parameter $n$, and cannot be independently determined within the framework of the model. However, for any reasonable values of the parameters it seems impossible to achieve consistency with the data (rms<25%), if the NS radius is too small ($R<1.7\ R_g$) or too large ($R>4\ R_g$).

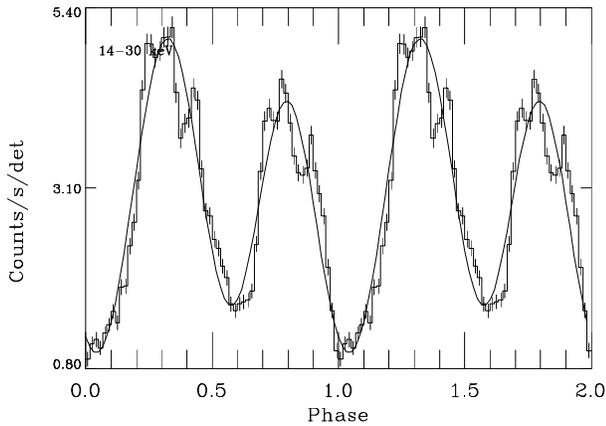

Fig. 7. Approximation of the experimental data by the model of the symmetrical polar caps with n=2.5, β=50°, γ=74.5°, Δϕ=10°, α=6°, Δβ=20°, ϕ$_0$=116°.6.

## CONCLUSIONS

During the observations of the X-ray pulsar GX 301-2 by the HEXE instrument onboard Mir-Kvant the essential changes in the object luminosity, connected with its passage near the optical companion were found. It was detected that at the brightness maximum the spectrum is softer than that in the low-luminosity state.

Interpretation of the pulse profiles within the framework of a model, which takes into account effects of the general theory of relativity and beaming of the emission, has shown, that the magnetic field of the object can not be qualitatively described by a dipole approximation, i.e. a quadrupole or other components should be present. The model shows also, that the magnetic axis is inclined to the axis of rotation by an angle ~40-70°, and the object is observed at ~75-85° with respect to the axis of rotation.

Table 5[a].
The pulse profile approximation by the model with two polar caps

| | | | Symmetrical cap | | | | | | | | Cap with angular width 120° | | | | |
|---|---|---|---|---|---|---|---|---|---|---|---|---|---|---|---|
| β[b] | Δβ[c] | Γ[d] | χ²/rms[e] | β | Δβ | Γ | χ²/rms | β | Δβ | Γ | χ²/rms | β | Δβ | Γ | χ² |
| | | | | | | | n =3 [f] | | | | | | | | |
| 35 | 0 | 87 | 823/13.6 | 45 | 0 | 86 | 859/15.0 | 45 | 0 | 86 | 923/14.5 | 60 | 0 | 86 | 894/20.6 |
| | 10 | 82 | 746/12.3 | | 10 | 81 | 896/16.1 | | 10 | 83 | 910/14.3 | | 10 | 80 | 882/21.7 |
| | 20 | 77 | 758/12.2 | | 20 | 76 | 883/15.4 | | 20 | 77 | 900/14.1 | | 20 | 76 | 800/21.1 |
| 40 | 0 | 87 | 759/12.7 | 85 | 0 | 43 | 887/14.0 | 50 | 0 | 85 | 867/15.2 | 65 | 0 | 86 | 908/24.8 |
| | 10 | 82 | 773/12.9 | | | | | | 10 | 82 | 852/15.8 | | 10 | 79 | 920/25.7 |
| | 20 | 76 | 785/12.9 | | | | | | 20 | 76 | 829/15.0 | | 20 | 75 | 947/25.1 |
| | | | | | | | | 55 | 0 | 85 | 868/17.3 | | | | |
| | | | | | | | | | 10 | 80 | 865/17.9 | | | | |
| | | | | | | | | | 20 | 77 | 769/17.5 | | | | |
| | | | | | | | n = 2.5 | | | | | | | | |
| 40 | 0 | 86 | 912/14.7 | 60 | 0 | 83 | 878/15.8 | 55 | 0 | 85 | 886/14.3 | 70 | 0 | 82 | 753/16.0 |
| | 10 | 81 | 820/13.4 | | 10 | 77 | 859/15.6 | | 10 | 80 | 892/14.1 | | 10 | 81 | 779/17.0 |
| | 20 | 76 | 811/13.3 | | 20 | 70 | 783/13.4 | | 20 | 76 | 797/13.1 | | 20 | 75 | 876/17.6 |
| 45 | 0 | 86 | 788/13.0 | 65 | 0 | 82 | 954/18.2 | 60 | 0 | 86 | 874/14.4 | 75 | 0 | 85 | 857/19.4 |
| | 10 | 80 | 749/12.4 | | 10 | 75 | 900/16.6 | | 10 | 79 | 776/13.6 | | 10 | 81 | 966/21.4 |
| | 20 | 75 | 748/12.2 | | 20 | 67 | 788/13.6 | | 20 | 76 | 705/13.1 | 85 | 0 | 47 | 890/13.4 |
| 50 | 0 | 85 | 758/12.7 | 70 | 10 | 69 | 917/16.5 | 65 | 0 | 84 | 766/14.7 | | | | |
| | 10 | 80 | 752/12.6 | 75 | 10 | 62 | 833/14.3 | | 10 | 79 | 696/14.3 | | | | |
| | 20 | 74 | 742/12.3 | 80 | 0 | 65 | 983/18.0 | | 20 | 75 | 714/14.2 | | | | |
| 55 | 0 | 84 | 802/13.8 | 85 | 0 | 49 | 758/12.6 | | | | | | | | |
| | 10 | 79 | 795/13.9 | | | | | | | | | | | | |
| | 20 | 73 | 760/12.9 | | | | | | | | | | | | |
| | | | | | | | n = 2 | | | | | | | | |
| 50 | 0 | 84 | 974/15.3 | 65 | 0 | 82 | 753/12.6 | 70 | 0 | 84 | 982/14.9 | | | | |
| | 10 | 79 | 917/14.8 | | 10 | 76 | 769/12.7 | | 10 | 81 | 922/14.3 | | | | |
| | 20 | 75 | 954/14.9 | | 20 | 68 | 834/13.5 | | 20 | 74 | 862/13.9 | | | | |
| 55 | 0 | 83 | 847/13.9 | 70 | 0 | 72 | 749/12.6 | 75 | 0 | 85 | 884/14.1 | | | | |
| | 10 | 78 | 827/13.5 | | 10 | 72 | 758/12.7 | | 10 | 80 | 860/14.3 | | | | |
| | 20 | 74 | 876/14.0 | | 20 | 67 | 884/13.8 | 80 | 0 | 85 | 878/14.6 | | | | |
| 60 | 0 | 83 | 780/13.0 | 75 | 0 | 75 | 747/12.6 | 85 | 0 | 80 | 878/14.6 | | | | |
| | 10 | 77 | 780/12.9 | | 10 | 67 | 787/13.0 | | | | | | | | |
| | 20 | 71 | 834/13.6 | 80 | | 68 | 748/12.6 | | | | | | | | |
| | | | | 85 | | 60 | 833/13.4 | | | | | | | | |

[a] In all cases the initial phase is $\phi_0 \sim 116°$; the azimutahal shift of the second pole in respect to the diametrically opposit position is $\Delta\phi \sim -10°$; all angles are given in degrees
[b] Angle between the magnetic axis and the axis of rotation
[c] Displacement of the second pole in direction of the disk plane
[d] Angle between the direction to observer and axis of the pulsar rotation
[e] Number of degress of freedom is 55; typical errorbar of *rms* is 0.5-1%
[f] $n$ determines the beaming factor (see eq. 3)

# References

Alexandrovich N.L., Arefiev V.A., Borozdin K.N., Kaniovsky A.S., Sunyaev R.A. // PAZh, 1994, V. 20, P. 660.
Basko M.M., Sunyaev R.A. // Astron. Astrophys., 1975, V. 42, P. 311.
Basko M.M., Sunyaev R.A. // MNRAS, 1976, V. 175, P. 395.
Vidal N.V. // Asprophys. J., 1973, V. 186, L. 81.
Jones C.A., Chetin T., Liller W. // Asprophys. J., 1974, V. 190, L. 1.
Ghosh P., Lamb F.K. // Asprophys. J., 1978, V. 223, L. 83.
Ghosh P., Lamb F.K. // Asprophys. J., 1979à, V. 232, P. 259.
Ghosh P., Lamb F.K. // Asprophys. J., 1979á, V. 234, P. 296.
Kaper L., Lamers H.J.G.L.M., Ruymaekers E. // Astron. Astrophys, 1995, V. 300, P. 446.
Koh D.T., Bildsten L., Chakrabarty D., Nelson R.W., Prince T.A., Vaughan B.A., Finger M.H., Wilson R.B., Rubin B.C. // Astrophys. J, 1997, V. 479, P. 933.
Krasnobayev K.V., Sunyaev R.A. // PAZh, 1977, T. 3, C. 124.
Leahy D.A., Darbro W., Elsner R.F., Weisskopf M.C., Kanh S., Sutherland P.G., Grindlay J.E. // Astrophys. J., 1983, V. 266, P. 160.
Leahy D.A., Matsuoka M. // Astrophys. J., 1988, V. 355, P. 627.
Lutovinov A.A., Grebenev S.A., Sunyaev R.A., Pavlinsky M.N. // PAZh 1994, Ò. 20, Ñ. 631.
Lampton M., Margon B., Bowyer S. // Astrophys. J., 1976, V. 208, P. 177.
Meszaros P., Nagel. W. // Astrophys. J., 1985, V. 299, P. 138.
Parkes G.E., Culhane J.L., Mason K.O., Murdin P.G. // MNRAS, 1980, V. 191, P. 547.
Pechenick K.R., Ftaclas C., Cohen J.M. // Astrophys. J., 1983, V. 274, P. 846.
Pravdo S.H., Day C.S.R., Angelini L., Harmon B.A., Yoshida A.A., Saraswat P. // Astrophys. J., 1995, V. 454, P. 872.
Reppin C., Pietch W., Trumper J., Kendziorra E., Staubert R. // Non-Thermal and very high Temperature Phenomena in X-Ray Astronomy, ESLAB Symp. 17 (ed. Perolla G.C, Salvati M.), Rome: Instituto Astronomico., 1983, P. 279.
Riffert H., Meszaros P. // Astrophys. J., 1987., V. 325., P.207.
Riffert H., Nollert H.-P., Kraus U., Ruder H. // Astrophys. J., 1993, V. 406, P. 185.
Rothschild R. E., Song Y. // Astrophys. J., 1986, V. 315, P. 154.
Sato N., Nagase F., Kawai N., Kelley R.L., Rappaport S., White N.E. // Astrophys. J., 1986, V. 304, P. 241.
Stevens I.R. // MNRAS, 1988, V. 232, P. 199.
White N.E., Swank J.H., Holt S.S. // Asptophys. J, 1983, V. 270, P. 711.
White N.E., Swank J.H. // Astrophys. J., 1984, V. 287, P. 856.
Haberl F. // Astrophys. J., 1991, V. 376, P. 245.
Chichkov M.A., Sunyaev R.A., Lapshov I. Yu. // PAZh, 1995, V. 21, P. 491.

Table 6
Restrictions on the NS parameters

| n | $\beta_{simm}$[1], deg | $\beta_{non.simm}$, deg[2] |
|---|---|---|
| 3 | 35-40 | 55-60 |
| 2.5 | 45-50 | 60-70 |
| 2 | 65-75 | >75 |

[1] The angle between the magnetic and the rotation axes
[2] Cap with azimuthal length 120°